\documentclass[twocolumn,trackchanges]{aastex7}

\usepackage{graphicx}	
\usepackage{amsmath}	

\def\rsh{r_{\rm sh}}

\def\vou{v_{r{\rm ou}}}
\def\rou{r_{\rm ou}}
\def\theou{\Theta_{\rm ou}}
\def\lou{\lambda_{\rm ou}}

\def\mbh{M_{\rm BH}}
\def\rin{r_{\rm in}}
\def\rg{r_{\rm g}}
\def\tg{t_{\rm g}}
\def\rhou{\rho_{\rm ou}}

\def\msol{M_\odot}
\def\medd{\dot{M}_{\rm Edd}}
\def\mdoti{\dot{M}_{\rm in}}
\def\mdoto{\dot{M}_{\rm acc}}
\def\mou{\dot{M}_{\rm ou}}
\def\tin{t_{\rm ini}}
\def\lbol{L_{\rm bol}}
\def\me{m_{\rm e}}
\def\mp{m_{\rm p}}
\def\ne{n_{\rm e}}
\def\np{n_{\rm p}}
\def\lsim{\lower.5ex\hbox{$\; \buildrel < \over \sim \;$}}
\def\gsim{\lower.5ex\hbox{$\; \buildrel > \over \sim \;$}}
\def \simeq{\lower.3ex\hbox{$\; \buildrel \sim \over - \;$}}

\begin{document}

\title{Dynamical properties of oscillating, viscous, transonic accretion disks around black holes}

\author[orcid=0000-0002-9851-8064,sname='Debnath']{Sanjit Debnath}
\affiliation{Aryabhatta Research Institute of Observational Sciences (ARIES), Manora Peak, Nainital 263001, India}
\affiliation{Department of Applied Physics, Mahatma Jyotiba Phule Rohilkhand University, Bareilly, Uttar Pradesh, 243006, India}
\email[show]{sdebnath@aries.res.in}  

\author[orcid=0000-0002-2133-9324,sname='Chattopadhyay']{Indranil Chattopadhyay} 
\affiliation{Aryabhatta Research Institute of Observational Sciences (ARIES), Manora Peak, Nainital 263001, India}
\email[show]{indra@aries.res.in}

\author[orcid=0000-0002-9036-681X,sname=Joshi]{Raj Kishor Joshi}
\affiliation{Nicolaus Copernicus Astronomical Center, Polish Academy of Sciences, Bartycka 18, PL-00-716 Warsaw, Poland}
\email{rjoshi@camk.edu.pl}

\author[orcid=0000-0001-9094-0335,sname=Laurent]{Philippe Laurent}
\affiliation{IRFU / Service d'Astrophysique, Bat. 709 Orme des Merisiers, CEA Saclay, 91191 Gif-sur-Yvette, Cedex France}
\email{fakeemail4@google.com}

\author[orcid=0009-0002-7498-6899,sname=Tripathi]{Priyesh Kumar Tripathi}
\affiliation{Aryabhatta Research Institute of Observational Sciences (ARIES), Manora Peak, Nainital 263001, India}
\affiliation{Department of Applied Physics, Mahatma Jyotiba Phule Rohilkhand University, Bareilly, Uttar Pradesh, 243006, India}
\email{fakeemail5@google.com}

\author[]{M. Saleem Khan}
\affiliation{Department of Applied Physics, Mahatma Jyotiba Phule Rohilkhand University, Bareilly, Uttar Pradesh, 243006, India}
\email{fakeemail6@google.com}

\begin{abstract}
We investigate the time evolution of sub-Keplerian transonic accretion flow onto a non-rotating black hole using axisymmetric viscous hydrodynamic simulations. We simulate the accretion flow using boundary values from semi-analytical analysis and set up three different models. Two of the models do not predict accretion shocks from the semi-analytic analysis, while one of them does. 
We also consider radiative cooling along with viscosity in the simulation. Our two-dimensional simulation deviated from the one-dimensional semi-analytical solution and admitted shocks in all three models. Viscous dissipation tends to push the shock front outward, and radiative cooling will push it in. Additionally, gravity is attractive. Depending on the competing strengths of all three processes, it may trigger shock oscillation. Different rates of angular-momentum transport in various layers may trigger eddies, which will enhance the shock oscillation. We show that any simple power law cannot approximate these solutions. We find that hot and higher angular-momentum flow requires higher viscosity to produce oscillatory shocks. From the temporal variation of the luminosity, shock oscillations generate QPOs in the range of sub-Hertz to a few Hertz frequencies if a ten solar mass black hole is assumed. 
\end{abstract}
\keywords{\uat{Black hole physics}{159} --- \uat{Accretion}{14} --- \uat{Hydrodynamics}{1963} --- \uat{Shocks}{2086} --- \uat{High energy astrophysics}{739}}

\section{Introduction}
Disk accretion is likely the primary mechanism responsible for the production of energetic radiation from active galactic nuclei and microquasars. Following the foundational studies by \cite{1952MNRAS.112..195B}, \cite{1972A&A....21....1P}, and \cite{1973A&A....24..337S}, numerous research papers have explored disk accretion onto gravitating objects. \cite{1973A&A....24..337S} assumed an accretion disk with Keplerian angular momentum distribution, which is geometrically thin and optically thick. They used a parametric viscosity prescription to remove the angular momentum outwards. This disk is known as the Keplerian disk (KD). KD is applicable for systems characterized by low mass accretion rates and can reproduce the thermal part of the continuum electromagnetic spectra, which is a modified blackbody. However, KD was unable to reproduce the power-law part of the spectrum. Moreover, a black hole X-ray binary (BHXRB) shows distinct spectral/temporal changes in various energy bands. These sources can exhibit radio jets and quasi-periodic oscillations (QPOs) in the power-law part of the continuum spectra, depending on their spectral states \citep[for details see][]{2004MNRAS.355.1105F,2019NewAR..8501524I,2024MNRAS.531.1149N}). 

The inability of the KD to explain the hard power-law tail in the spectral energy distribution (SED) of microquasars prompted a plethora of accretion-disk models in the quest to address various observational features as they were discovered. One of the most important accretion disk models that followed the KD was the thick disk model (or TD), also sometimes called the ‘Polish doughnut’ \citep{1976ApJ...207..962F,1980A&A....88...23P,1980ApJ...242..772A}. Unlike KD, TD retained the pressure gradient term but, similar to KD, ignored the advection term in the equations of motion. However promising it was, the lack of the advection term made the TD unstable \citep{1984MNRAS.208..721P}. 
The next very popular model included the advection term and was called
advection-dominated accretion flows (ADAF) \citep{1977ApJ...214..840I,1995ApJ...452..710N}, and for super Eddington accretion rates it was slim accretion flows \citep{1988ApJ...332..646A}. 

The theoretical model of accretion disk with a standing shock \citep{1987PASJ...39..309F, 1989ApJ...347..365C,2008ApJ...677L..93B,2009ApJ...702..649D, 2011IJMPD..20.1597C, 2013MNRAS.430..386K, 2014MNRAS.443.3444K} offers a self-consistent framework for the formation of a corona --- a reservoir of hot electrons and is the Comptonizing cloud for soft photons \citep{1995ApJ...455..623C,2017MNRAS.465.3902C}. 
Several stability analyses \citep{1992MNRAS.259..259N, 1994PASJ...46..257N} have examined the shock in the accretion disk and found that the outer shock is stable. 
Many numerical simulations studied the stability properties of shocks in transonic, inviscid accretion disks \citep{1994ApJ...425..161M, 1996ApJ...457..805M, 1996ApJ...470..460M, 1995ApJ...452..364R,2010MNRAS.403..516G,2012MNRAS.425.2413O, 2012ApJ...758..114G,2014MNRAS.437.1329G,2017MNRAS.472.4327S,2017MNRAS.472..542K,2019MNRAS.482.3636K, 2020ApJ...904...21P, 2022MNRAS.514.5074O, 2023A&A...678A.141O,2023ApJ...946L..42K,2025ApJ...990...12M}.

Since viscosity transports angular momentum and the accretion shock is rotation-mediated, simulations of transonic viscous disks showed persistent shock oscillations \citep{1998MNRAS.299..799L, 2011ApJ...728..142L,2012MNRAS.421..666G,2013MNRAS.430.2836G,2014MNRAS.442..251D,2015MNRAS.453..147O,2015MNRAS.448.3221G,2016ApJ...831...33L}. The shock oscillation frequencies matched with those of the quasi-periodic oscillations (QPOs) observed in BHXRB. 
However, these simulations use supersonic outer-boundary conditions, which are unrealistic. If an accretion solution has no shock transition, viscosity becomes ineffective in a simulation run with supersonic injection since the flow remains supersonic throughout the domain. If there's a shock, then a supersonic injection implies that the accretion is supersonic from the injection point to the shock, and subsonic in the post-shock region. It again becomes supersonic very close to the Black hole. In the pre-shock flow, viscosity is ineffective, and there's a shallow gradient in angular momentum distribution. On the other hand, in the post-shock disk, viscosity becomes significant and angular momentum piles up. Such an imbalance across the shock may trigger spurious oscillation.
Furthermore, a fixed adiabatic index ($\Gamma$) equation of state (EoS) for the gas was used. An accretion disk can extend up to several thousand Schwarzschild radii, and the flow temperature varies by a few orders of magnitude. Over such a broad temperature range, using a fixed $\Gamma$ EoS is unrealistic, as $\Gamma$ is inherently temperature-dependent. \cite{2008AIPC.1053..353C, 2009ApJ...694..492C} introduced a variable $\Gamma$ equation of state (EoS) known as the CR EoS. This EoS has been implemented in simulations and addressed a variety of astrophysical problems \citep{2013ASInC...9...13C, 2022MNRAS.509...85J, 2022ApJ...933...75J, 2023ApJ...948...13J, 2024ApJ...971...13J,2025ApJ...979...61T}. 

\cite{2024MNRAS.528.3964D} implemented the CR EoS in a one-dimensional code and regenerated all the steady state accretion solutions \citep[similar to][]{2014MNRAS.437.2992K} onto a black hole (BH). With subsonic injection, there can be comparable angular momentum distribution across the shock, although angular momentum piling in the post-shock disk still occurs. However, the imbalance in angular momentum distribution across the shock is not too high, which would mitigate instabilities triggered by boundary conditions. They also demonstrated viscosity-driven shock oscillations and their relation to QPOs in micro-quasars. However, one-dimensional simulations are limited.
In this paper, we study two-dimensional (2D), transonic, viscous accretion flows with consistent cooling. 
We want to study how viscosity affects the dynamics of the accretion disk. Unlike our previous one-dimensional study, we have included radiative cooling in the equations of motion in this paper. Can vortices develop at higher viscosity? How does viscosity influence outflow properties?
How does cooling affect the disk-outflow physics? What is the impact of injected angular momentum on disk dynamics? By varying the combination of viscosity and angular momentum in our simulations, we address these questions in this work.

The structure of this paper is as follows. We briefly discuss the model equations in the section \ref{sec:model_equations}. Section \ref{sec:Numerical_setup} describes the numerical simulation method, including the computational grid, boundary conditions, and initial setup. In Section \ref{sec:result}, we present the results of our numerical simulations. Finally, in Section \ref{sec:summary}, we summarize the findings and relevance of our numerical simulation results.
\section{MODEL EQUATIONS} \label{sec:model_equations}
\subsection{Hydrodynamic equations} 
We solve the conserved fluid equations in the spherical coordinate system $(r, \theta, \phi)$. Assuming the axisymmetry ($\partial/\partial\phi=0$), these equations can be expressed in the conservative form as:
\begin{equation}
 \frac{\partial \mathbf{q}}{\partial t} + \frac{1}{r^2} \frac{\partial (r^2\mathbf{F^r})}{\partial r} +\frac{1}{r \sin\theta} \frac{\partial(\sin\theta \mathbf{F^\theta})}{\partial \theta}  = \mathbf{S}.
 \label{eq:conserve}
\end{equation}
The conserved variables are $\mathbf{q}$s, while the primitive variables are $\mathbf{w}$s,
\begin{equation}
\mathbf{q}= 
\begin{bmatrix}
\rho \\
 M_r \\
 M_{\theta} \\
  M_\phi \\
E \\
\end{bmatrix}
=
\begin{bmatrix}
\rho \\
 \rho v_r \\
\rho v_{\theta} \\
 \rho v_{\phi} \\
\rho v^2/2 +e \\
\end{bmatrix};
~~ \mathbf{w}= 
\begin{bmatrix}
\rho \\
 v_r \\
 v_{\theta} \\
  v_{\phi} \\
 p \\
\end{bmatrix} 
\label{eq:state_primitv}
\end{equation}
where, the fluxes corresponding to the $\mathbf{q}$s are given as
\begin{equation}
\mathbf{F^r}= 
\begin{bmatrix}
\rho v_r \\
v_r M_r+p \\
v_r M_{\theta} \\
v_r M_\phi \\
(E+p)v_r\\
\end{bmatrix} ,
~~ \mathbf{F^\theta}= 
\begin{bmatrix}
\rho v_\theta \\
 v_\theta  M_r \\
 v_{\theta} M_{\theta}+p \\
  v_{\theta} M_{\phi}\\
 (E+p)v_\theta \\
\end{bmatrix},
\label{eq:flux}
\end{equation}
and the source terms of the equations of motion are given as
\begin{equation}
\mathbf{S}= 
\begin{bmatrix}
0 \\
\frac{\rho v^2_{\phi}}{r}+\frac{\rho v^2_{\theta}}{r} -\frac{\rho G M_{BH}}{(r-r_g)^2}+\frac{2p}{r} \\
-\frac{\rho v_r v_{\theta}}{r}+\frac{\rho v^2_{\phi} \cot\theta}{r}+\frac{p \cot\theta}{r}\\
-\frac{\rho v_r v_\phi}{r}-\frac{\rho v_\theta v_\phi \cot\theta}{r}+ S_\lambda \\
 -\frac{G M_{BH} \rho v_r}{(r-r_g)^2}+S_E -S_Q \\
\end{bmatrix}.
\label{eq:Smatrix}
\end{equation}
The set of equations represented by Eq.\eqref{eq:conserve} consists of the continuity equation, the three momentum equations, and the energy equation. The strong gravity is mimicked by Paczy{\'n}sky-Wiita potential \citep{1980A&A....88...23P}. Here, $\rho$ denotes the fluid rest mass density, $M_r=\rho v_r$, $M_{\theta}=\rho v_\theta$, $M_{\phi}=\rho v_\phi$ represent the momentum density in the $r$ and $\theta$, and $\phi$ directions, respectively. $E=\rho v^2/2+e$ represents the total energy density, including both the internal and kinetic energy density. Here, $p$ is the pressure, and $e$ is the sum of thermal and rest mass energy density of the gas. The velocity components $v_r$ and $v_\theta$ correspond to the radial and polar directions, while $\lambda=rv_{\phi}$ is the specific angular momentum, with $v_{\phi}$ being the azimuthal component of the velocity. Additionally, the total velocity squared is expressed as $v^2=v_r^2+v_\theta^2+v_\phi^2$.
We consider only the $r-\phi$ component($W_{r\phi}$) of the viscous stress. 
The viscous stress is given by,
\begin{equation}
W_{r\phi} = \eta_v r \frac{d\Omega}{dr}.
\label{eq:stress}
\end{equation}
where $\Omega$ represents the angular velocity. So, using $W_{r\phi}$ component, $S_{\lambda}$ and $S_E$ are given as,
\begin{equation}
S_{\lambda}= {\frac{1}{r^3}\frac{\partial}{\partial r}\left(r^3 W_{r\phi}\right);} ~
~S_E= \frac{1}{r^2}\frac{\partial ( r^2 v_{\phi} W_{r\phi})}{\partial r},
\label{eq:sesl}
\end{equation}
$\eta_v = \rho \nu$ is the dynamic viscosity coefficient, {$\nu= (\alpha p)/(\rho \Omega_k)$} denotes the kinematic viscosity. In this context,  $\alpha$ refers to the Shakura–Sunyaev viscosity parameter. $\Omega_k$ is the local Keplerian angular velocity and can be expressed as,
\begin{equation}
\Omega^2_k = \frac{1}{r} \frac{d \Phi}{d r}.
\label{eq:omegak}
\end{equation}
$S_Q$ is the radiative cooling contribution. In this work, we have considered both bremsstrahlung and synchrotron radiation. The emissivity resulting from bremsstrahlung (measured in ${\mathrm{erg}~\mathrm{cm}^{-3}~\mathrm{s}^{-1}}$) is described by \cite{1973blho.conf..343N} and
given as,
\begin{equation}
Q_{\rm br}=1.4\times10^{-27} n^2_e\sqrt{T_e}\left(1+4.4\times10^{-10}T_e\right).
 \label{eq:bremm}
\end{equation}
and the emissivity due to the synchrotron (in ${\mathrm{erg}~ {\rm cm}^{-3}~ {\rm s}^{-1}}$) is given by \cite{1983JBAA...93R.276S},
\begin{equation}
Q_{\rm syn}=\frac{16}{3} \frac{q_e^2}{c} \left( \frac{q_eB}{m_e c} \right)^2 \Theta^2_{e} n_e.
 \label{eq:synchro}
\end{equation}
$T_{\rm e}$ and $n_{\rm e}$ represent the electron temperature and number density, respectively. $\Theta_e=k_b T_e/m_e c^2$ is the dimensionless electron temperature, estimated following \citep{2014MNRAS.437.2992K,2025ApJ...979...61T}. $q_e$ is the charge of the electron. $B$ represents the random magnetic field, and $B^2/(8 \pi)=\beta p$, where the proportionality parameter $\beta$ is takes as 0.3.  
So, $S_Q=Q_{\rm br}+Q_{\rm syn}$.
\subsection{Equation of state} 
We adopt a variable adiabatic index EoS for multispecies fluids \citep{2009ApJ...694..492C}, known as the CR EoS. 
More recently, \cite{2021MNRAS.502.5227J} presented the CR EoS by redefining the temperature variable as $\Theta=p/(\rho c^2)$, and is given by

\begin{equation}
 e = \rho c^2 f
 \label{eq:eos}
\end{equation}
where, $f$ is given as 
\begin{equation}
 f = 1 + (2-\xi)\Theta\left[\frac{9\Theta+6/\tau}{6\Theta+8/\tau}\right]+\xi\Theta\left[\frac{9\Theta+6/\tau\eta}{6\Theta+8/\tau\eta}\right].
 \label{eq:f}
 \end{equation}
Here, $\rho=\sum n_{\rm i} m_{\rm i}=\ne \me(2-\xi+\xi/\eta)$, where $\xi =\np/\ne$, $\eta= \me /\mp$, and $n_p$, $\ne$, $\mp$, and $\me$ represent the proton number density, electron number density, proton rest mass, and electron rest mass, respectively. Here $\tau = 2-\xi + \xi/\eta$. In this study, we assume $\xi=1$ or electron-proton flow. The expression for specific enthalpy and sound speed is also given as,
\begin{equation} 
 h=(e+p)/\rho=(f+\Theta)c^2;~~c_s=\sqrt{\Gamma~\Theta}
 \label{eq:h}
\end{equation}
The adiabatic index is $\Gamma=1+1/N$ is related to polytropic index $N$, 
\begin{equation} 
\begin{aligned}
N=\rho\frac{\partial h}{\partial p}-1=\frac{\partial f}{\partial\Theta}
     =6\left[(2-\xi)\frac{9\Theta^2+24\Theta/\tau+8/\tau^2}{(6\Theta+8/\tau)^2}\right] \\
+6\xi\left[\frac{9\Theta^2+24\Theta/(\eta\tau)+8/(\eta\tau)^2}{(6\Theta+8/(\eta\tau))^2}\right].
 \label{eq:N}
 \end{aligned}
\end{equation}


\begin{deluxetable*}{rllll}
\tablewidth{0pt}
\tablecaption{Details of the injection parameters: radial velocity ($\vou$), temperature ($\theou$), and specific angular momentum ($\lou$) at the outer boundary ($\rou$). 
\label{tab:injection_setup}}
\tablehead{
\colhead{Model} & \colhead{$\vou$} & \colhead{$\theou$} & \colhead{ $\lou$} & \colhead{$\rou$}
}
\startdata
L1 ($\epsilon=$1.0002)&   $-2.095\times 10^{-2} $    & $3.585\times 10^{-4} $ & 1.60 & 500 \\
L2 ($\epsilon=$1.0002)    &  $-2.090\times 10^{-2} $    & $3.598\times 10^{-4} $ & 1.72 & 500\\
L3 ($\epsilon=$1.0015)   &  $-0.895\times 10^{-2} $    & $8.479\times 10^{-4} $ & 1.90 & 500 \\
\enddata
\end{deluxetable*}
\begin{figure*}
	\begin{center}
       \includegraphics[width=\textwidth]{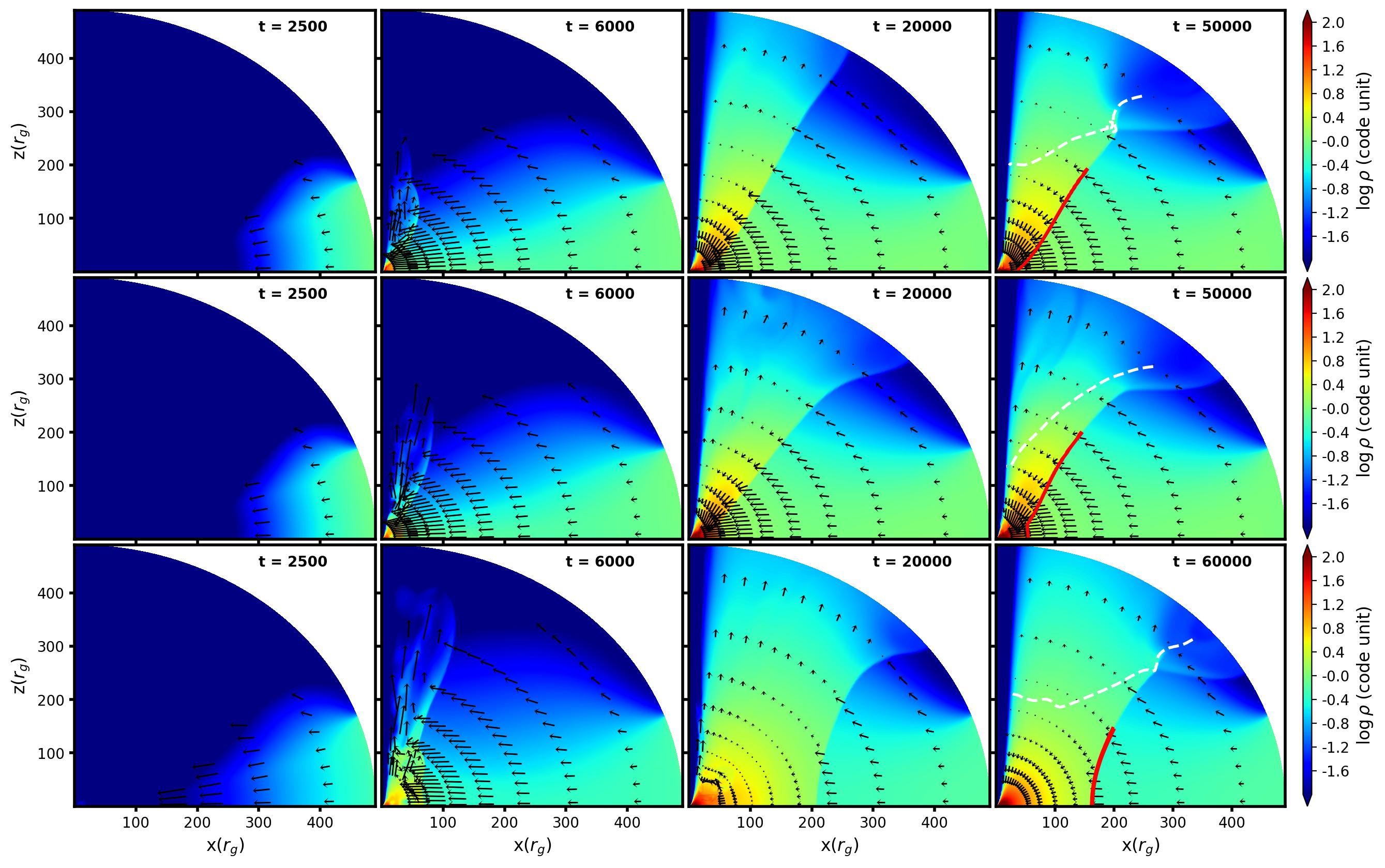}
        \caption{Initial Condition: Contours of logarithmic density (in code unit) overlaid with velocity vectors for the model L1 (First row), L2 (second row), and L3(third row). The times of snapshots are mentioned in the plots. The red and white dashed lines in the 4th column correspond to the shock surface and $v_r=0$, respectively. See text for details.}
        \label{fig:1}
        \end{center}
\end{figure*}

\section{NUMERICAL SET-UP} \label{sec:Numerical_setup}
\subsection{Code}
We numerically solve the conservative form of the fluid equations presented in Eq.\eqref{eq:conserve}. The unit of length is the Schwarzschild radius $\rg=2G\mbh/c^2$,
the unit of speed is the speed of light $c$, therefore, the unit of time is $\tg=\rg/c$ (where $G$ is the gravitational constant, $\mbh$ is the black hole mass). In this system, the unit of specific angular momentum is $\rg c$. 
Our code employs the finite volume method to ensure local conservation of fluid quantities within the computational grid. Broadly speaking, these methods consist of three main steps: first, an interpolation routine; second, solving Riemann problems at the zone edges; and finally, the time evolution.
The interpolation routine constructs left and right interpolated states of primitive variables around each cell boundary using the Minmod limiter. 
After this reconstruction, a Harten-Lax-van Leer (HLLC) Riemann solver \citep{doi:10.1137/1025002,Toro1994} is employed to calculate fluxes across cell boundaries. We use the second-order, two-stage, total variation diminishing (TVD) Runge-Kutta method for the time integration. 
For the parallel processing functionality, the code employs domain decomposition to distribute computations across multiple processors, with inter-processor communication managed using the Message Passing Interface (\texttt{MPI}) library.
\subsection{Numerical grid and boundary conditions} We use $(r, \theta)$ coordinates in the simulations. The simulation domain extends radially from $r_{\text{in}} = 1.8$ to $r_{\text{out}} = 500$ and spans the polar angle from $0$ to $\pi/2$. A uniform grid of 180 cells is used in the polar direction, while 642 logarithmic grid points are used in the radial direction, so the aspect
ratio $\Delta r/r \Delta \theta = 1$. Four ghost cells are implemented in each direction. Matter enters the computational domain through the outer boundary.
We apply a continuous inflow boundary condition at the inner boundary. 
The accretion rate is scaled in units of the Eddington accretion rate. Therefore, the unit of density is the injected density at the outer boundary
\begin{equation}
\rho_{\rm un}= \frac{1.44\times 10^{17}{\dot m}(\mbh/\msol)}{4\pi cos(\theta_0)\rhou \rou^2 v_r \rg^2 c},
\label{eq:den_unit}
\end{equation}
Here, $\dot{m}$ and $\msol$ are accretion rate in units of Eddington accretion rate ($\medd=1.44\times10^{17}
[\mbh/\msol)$gs$^{-1}$]) and the mass of the Sun, respectively. The matter is injected axisymmetrically in a cone of angle (co-latitude) between $\theta_0=7\pi/18$ to $\pi/2$ at $r=\rou$, assuming up-down symmetry. In this paper, we have considered only one accretion rate for all the models, i.e., $\dot{m}=0.3$.
The initial background material within the simulation domain has a density of $\rho_{\text{bg}} = 10^{-7}$. The initial pressure in the computational domain is computed using dimensionless temperature $\theou$. Axisymmetric boundary condition is enforced about the axis ($\theta = 0$), while equatorial symmetry is applied at the mid-plane ($\theta = \pi/2$). An outflow boundary condition is used at the outer radial boundary, except in the region $\theta_0=7\pi/18$ to $\pi/2$, where a fixed inflow boundary condition is imposed.

\subsection{Initial set-up}
The injection parameters of the three models L1, L2 \& L3 we consider in this paper are given in Table \ref{tab:injection_setup}. Here $\vou$ is the radial velocity, $\theou$ the temperature parameter, and $\lou$ the angular momentum at the radius $r=\rou$ adopted from the analytical solutions, and corresponds to specific energy $\epsilon$ and angular momentum $\lambda=\lou$ of the analytical solutions. The analytical solutions of L1 \& L2 are inviscid, and that of L3 is a viscous solution of $\alpha=0.001$ \citep[see,][for details]{2024MNRAS.528.3964D}, since the latter has higher angular momentum.  
The analytical solution of L1 \& L3 are shock-free
while L2 admits a stationary centrifugal barrier-mediated
shock at 17$\rg$. 

Using the injection parameters of Table \ref{tab:injection_setup}, we run the code until the time $t=t_{\rm in}$ when steady state is achieved. We treat this steady state as the initial condition for our study.
In all three models, radiative cooling is turned on from the start, and the accretion rate at the injection boundary is
\begin{equation}
    \mdoto=4\pi cos(\theta_{0})\rou^2 r^2_g \rhou \rho_{\rm un} \vou c = 0.3\medd.
\end{equation}
Therefore, $\mdoto$ is the mass supply for all the models considered in this paper.
The mass inflow rate into the BH is given by 
\begin{equation}
\mdoti=2 \pi \rin^2 \Sigma_j [\rho(\rin, \theta_j) v_r(\rin, \theta_j)\Delta \theta_j]
\label{eq:inflort}
\end{equation} 
Here $r_{\rm in}$ is the radius just outside the sink marking the BH, while $\rho_{\rm in}$ \& $v_{r{\rm in}}$
are the density and flow velocity at $r_{\rm in}$ and $j$ is the $\theta$ index.
The mass outflow rate computed is
\begin{equation}
\mou=2 \pi \rou^2 \Sigma_j [\rho(\rou, \theta_j)
v_r(\rou, \theta_j)\Delta \theta_j]
\label{eq:outflort}
\end{equation}
with the condition $v_r(\rou,\theta_j)>0$ \& greater than local sound speed. It must be further noted that we are simulating $\theta=0$ to $\pi/2$.
Therefore, ${\dot M}_{\rm in}$ and ${\dot M}_{\rm out}$ computed will be half of the total that is either entering the BH or leaving the system.

Figure \ref{fig:1} illustrates the time evolution of contours of log$(\rho)$ of models L1 (top row), L2 (middle row) \& L3 (bottom row) plotted at various time steps as they reach the steady state. Once the steady state is reached, it is treated as initial conditions for the three models when viscosity is turned on. The black arrows in the three panels represent velocity vectors for the three models. 
In models L1 \& L2, it takes a dynamical time of $t = 2500 \tg$ for the injected matter to reach $r=300\rg$ from the BH. The injected matter in L3 takes the same time to reach $r\sim 200\rg$ from the BH. 
By the time $t=6000\tg$ after the injection, the matter reaches the horizon for all three models. Accretion shocks form with collimated outflows emerging from the post-shock disk in all three models. 
In the 3$^{rd}$ column of Fig. \ref{fig:1}, the accretion and outflows are shown at $t=2\times 10^4 \rg$, in this time snap, the shock location continues to travel outward and has not reached its steady value. 
In column four, the steady state of models L1 \& L2 are shown at $t=50000\tg$, where the shock fronts settle at $\rsh=38\rg$ for L1 and $\rsh=53\rg$ for L2. 
For L3, the steady state is attained at $t=6\times10^4 \rg$, and the shock settles at around $\rsh=164 \rg$. It may be noted that the analytical solutions of L1 and L3 had no shock, while for L2 it was at 17$\rg$. Two-dimensional simulations do not match one-dimensional analytical solutions. Interestingly, the simulations predict shocks at parameter values that do not predict shocks from the analytical investigations. In 1D simulations, only one spatial dimension is dynamically important. For multi-dimensional simulations, degrees of freedom increase. In the present paper, matter not only accretes along the equatorial plane but also from off-equatorial regions (see the velocity vectors). And from these off-equatorial regions, matter can fall in or flow out, affecting the equatorial flow. The interaction of these two types of flows may satisfy the shock condition, even when purely 1D considerations did not. In the fourth column, the accretion shock surface and the outflow base (i.e., $v_r=0$) are indicated by the thick red line and the white dashed line, respectively. The time the three models reach steady state is referred to as the time to reach the initial state or $\tin$. So for L1 and L2, $\tin=5\times 10^4\tg$, but for L3, $\tin=6 \times 10^4\tg$.

\begin{figure}
	\begin{center} \includegraphics[width=3.3 in]{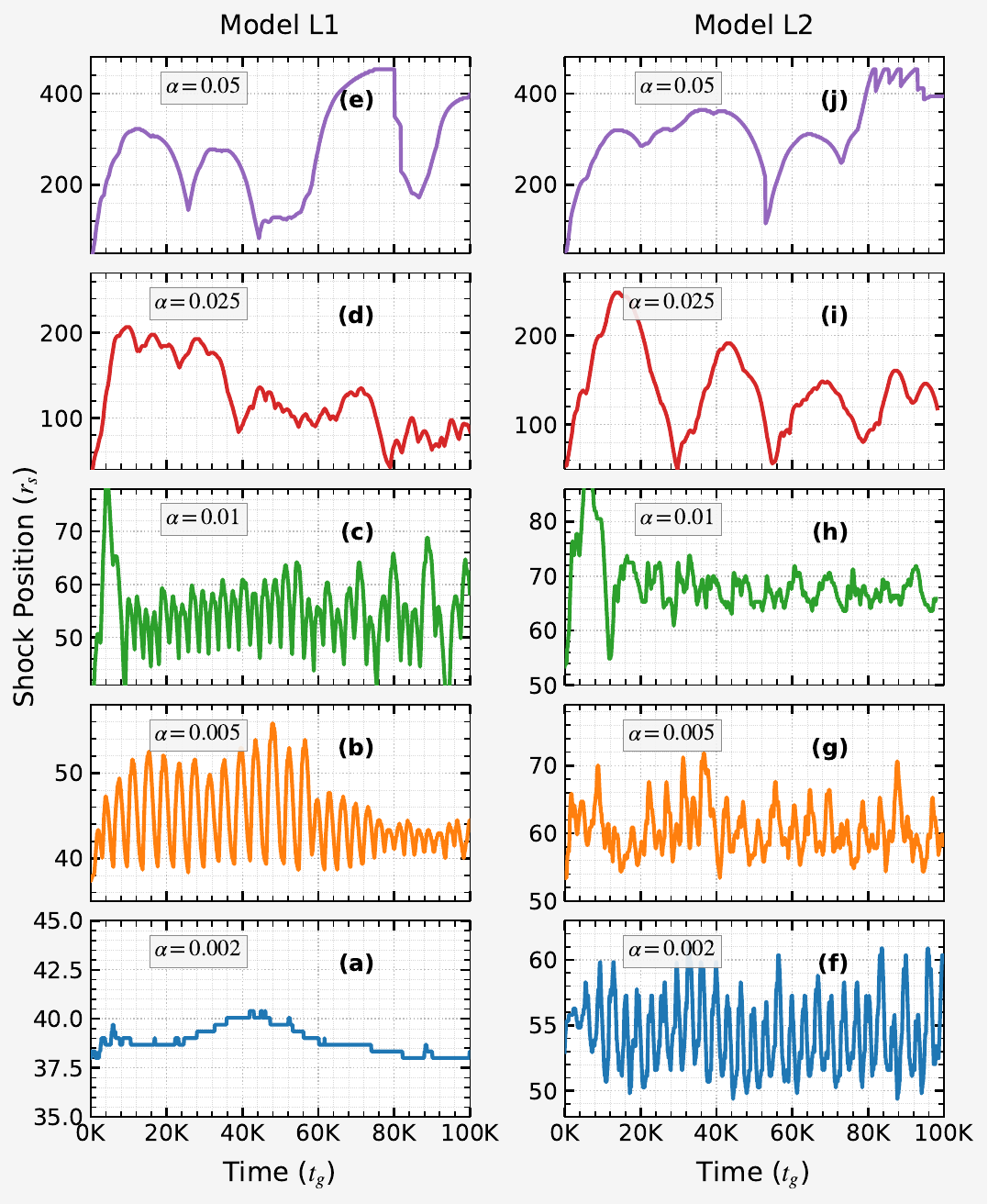}
        \caption{Time evolution of the outer shock position($r_s$) with time for different $\alpha$ = 0.002, 0.005, 0.01, 0.025, and 0.05 for the model L1 (left column) and the model L2 (right column). The time taken to achieve the initial condition for the simulation is rescaled as $\tin=0.0$ in the plots above.}
        \label{fig:2}
        \end{center}
\end{figure}

\section{RESULTS: OSCILLATING DISK}\label{sec:result}
We investigate the effect of viscosity where the initial condition is an inviscid steady state distribution in the presence of radiative cooling. 

\subsection{Effect of viscosity in the disk structure}\label{sec:visc_effect}

In this paper, we evolve simulations with injection parameters adopted from the analytical solutions until they reach a steady state. Then we turn on the viscosity and study the disk dynamics.
We have considered subsonic injection in all three models, which ensures non-negligible $\lambda$ transport between the injection boundary and the outer sonic point.

\begin{figure*}
	\begin{center}
       \includegraphics[width = \textwidth]{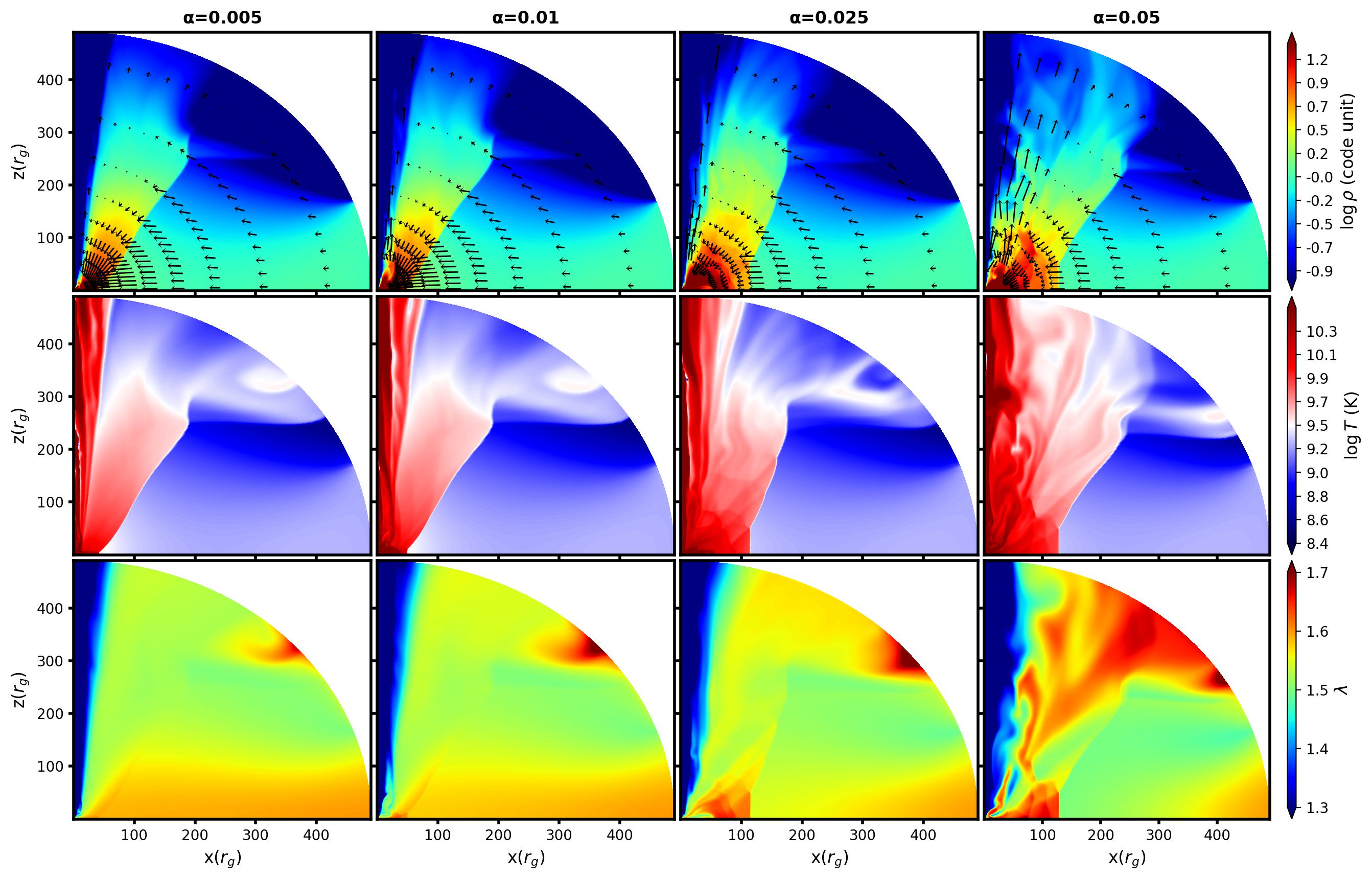}
        \caption{Distribution of logarithmic density($\rho$) in code unit overlaid with velocity vectors, logarithmic temperature(T) in Kelvin, and the angular momentum($\lambda$) at dynamical time t=50000$t_g$ for the model L1. The first, second, third, and last columns are for $\alpha = 0.005$, $\alpha = 0.01$, $\alpha = 0.025$, and $\alpha = 0.05$, respectively. See the text for more details.}
        \label{fig:3}
        \end{center}
\end{figure*}

Figure \ref{fig:2} shows the time series of the outer shock position ($\rsh$) for different $\alpha$ values of model L1 (panels a, b, c, d, e) and L2 (f, g, h, i, j). As viscosity transports angular momentum outwards, there is a piling up of angular momentum in the post-shock disk, since the post-shock disk is subsonic, except near the horizon. In the pre-shock flow, $\lambda$ distribution is flatter in the supersonic part, but increases outward beyond the outer sonic point.
This piling of $\lambda$ in the post-shock disk tends to push the shock front outward. However, where the shock will shift also depends on the gain in infall velocity and, thereby, the increase in inward ram pressure ($\rho v^2_r$) in the pre-shock disk. While the shock settles to a new steady position for $\alpha=0.002$ (Fig. \ref{fig:2}a) in model L1, for model L2 the shock undergoes regular oscillation for the same $\alpha$ (Fig. \ref{fig:2}f). In model L1, for $\alpha=0.005$, the shock initially exhibits regular oscillations. However, the amplitude gradually diminishes in time (Fig. \ref{fig:2}b). In contrast, for model L2 (Fig. \ref{fig:2}g), the shock exhibits persistent irregular oscillation when $\alpha$ = 0.005. As we increase the viscosity to $\alpha = 0.01$, the oscillation amplitude increases for both models L1 (Fig. \ref{fig:2}c) and L2 (Fig. \ref{fig:2}h). However, as expected, model L2 exhibits more irregular oscillations than model L1. Higher viscosity of $\alpha = 0.025$ (Figs. \ref{fig:2}d \& i) and $0.05$ (Fig. \ref{fig:2}e \& j) for both models L1 and L2, show increased irregularity in shock oscillation. Therefore, viscosity redistributes the angular momentum in a manner that triggers shock oscillation. Further increase in viscosity will degrade the oscillation, and the frequency of oscillation will decrease.

 \begin{figure*}
	\begin{center}
       \includegraphics[width = \textwidth]{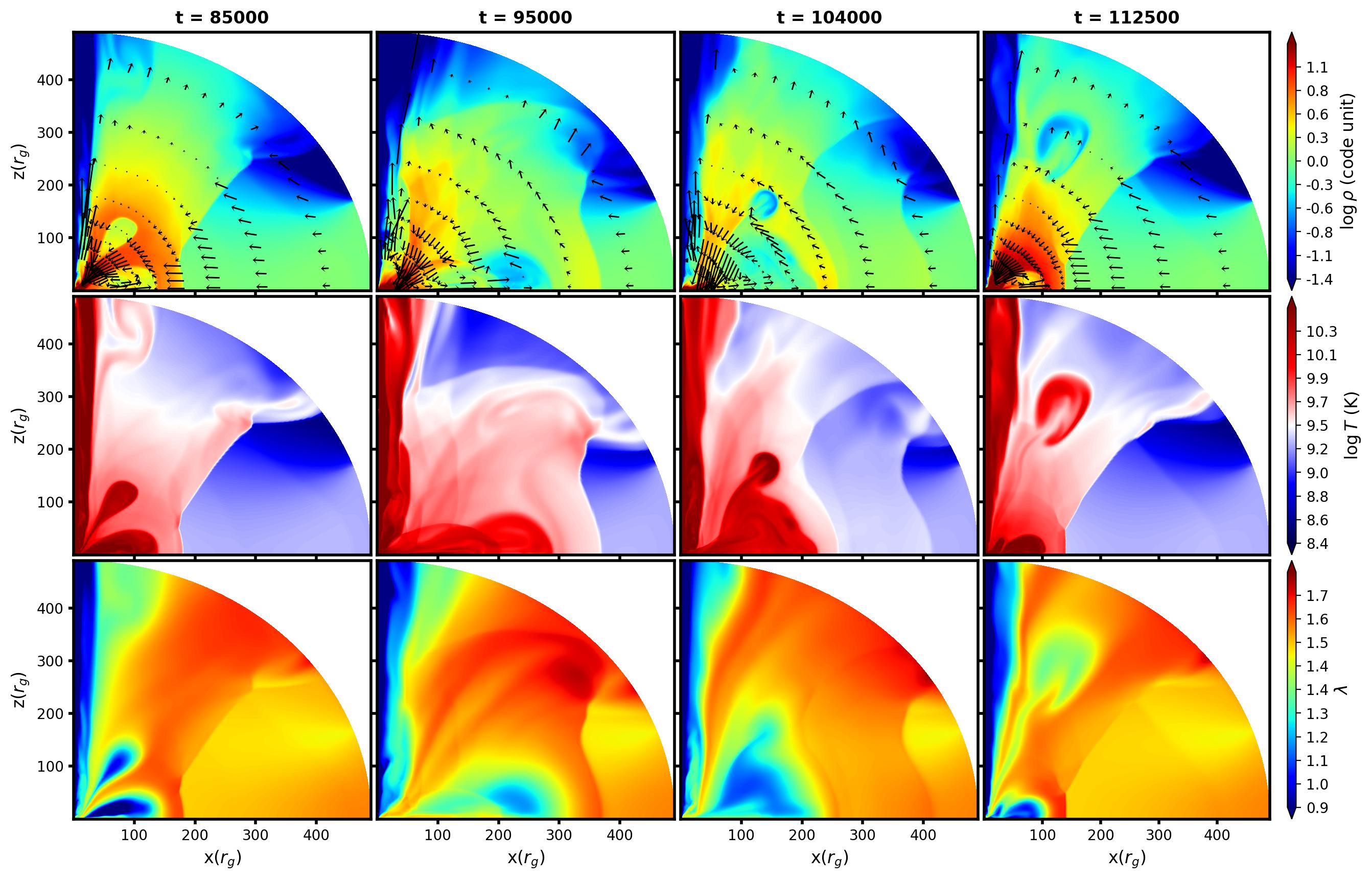}
        \caption{Contours of density and velocity vectors (arrows) in the first row, temperature (in K) in the second row, and angular momentum ($\lambda$) in the third row with the $\alpha = 0.05$ for the model L2. The first, second, third, and last columns are captured at times 85000$t_g$, 95000$t_g$, 104000$t_g$, and 112500$t_g$, respectively. See text for details.}
        \label{fig:4}
        \end{center}
\end{figure*}
In Fig. \ref{fig:3}, we present the density ($\rho$, top row in log scale), temperature (T, middle row in log scale), and angular momentum ($\lambda$, bottom row) for different values of $\alpha$ (mentioned on the panels) at time $t=\tin+50,000 t_g$ for the model L1. The first to fourth columns correspond to $\alpha = 0.005$, $\alpha = 0.01$, $\alpha = 0.025$, and $\alpha = 0.05$, respectively. The density jump from the pre-shock to post-shock disk increases by a factor of 4 to 5, and further inward, density increases up to an order of magnitude compared to the pre-shock value.
The temperature (Kelvin) distribution in the post-shock disk also shows an order of magnitude jump compared to the pre-shock disk temperatures. Angular momentum transport depends on the viscous stress (see Eq. \ref{eq:stress}), where the kinematic viscosity $\nu \propto \Theta$. Therefore, the angular momentum transport is more efficient in the post-shock disk. We plot the contours of $\lambda$ for different values of $\alpha$ in the lower panel. For lower values of $\alpha$, the shock front is close to the BH, and the $\lambda$ jump in the post-shock disk is faintly perceptible. For higher values, i.e., $\alpha=0.025,~0.05$, the angular momentum pile up in the post-shock disk is clearly visible. Additionally, the gradient of $\lambda$ is stronger in the pre-shock disk. The angular momentum distinctly decreases from the outer boundary towards the shock front.

Figure \ref{fig:4} illustrate the time evolution of log$\rho$, log$T$, and $\lambda$ distribution of model L2 for
$\alpha$ = 0.05. 
Hot bubble-like features are formed in the post-shock disk. The bubbles formed near the axis, leaving the system along the outflow.
Bubbles formed near the equatorial plane move outward, but are eventually pushed back due to the incoming injected material. 
Although the viscous tensor depends on the shear, the kinematic viscosity
$\nu ~ (\propto \Theta)$ is weak for supersonic flow. So, depending on shear, temperature, and the bulk velocity of the flow, viscous transport of angular momentum will be different near the equatorial plane and away from it. 
In the first row of Fig. \ref{fig:4}, contours of log($\rho$) is plotted at $t = 85000t_g,~95000\tg,~104000\tg,~112500\tg$ (marked on the figure). The corresponding contours of log($T$) and $\lambda$ are plotted in the middle and the bottom row, respectively. The regions where $\lambda$ transport is most efficient (low values) are also hotter and less dense. The inflow velocity in these regions is less than the surrounding regions, which, in time, grows into a strong outward push (velocity vectors are directed outwards). While in the surrounding regions, matter falls at a faster rate. This causes the formation of small eddies. In time, these eddies grow and push the outer shock front. The formation and dynamics of bubbles may be the reason behind the formation of multiple shock fronts. The outer two shock fronts on the equatorial plane merge at $t=112500$. Thereafter, the outward push weakens, and due to the ram pressure of the continuously injected matter, the shock front is pushed inward. This cycle continues, giving rise to oscillations. 

\begin{figure*}
	\begin{center}
    \includegraphics[width=\textwidth]{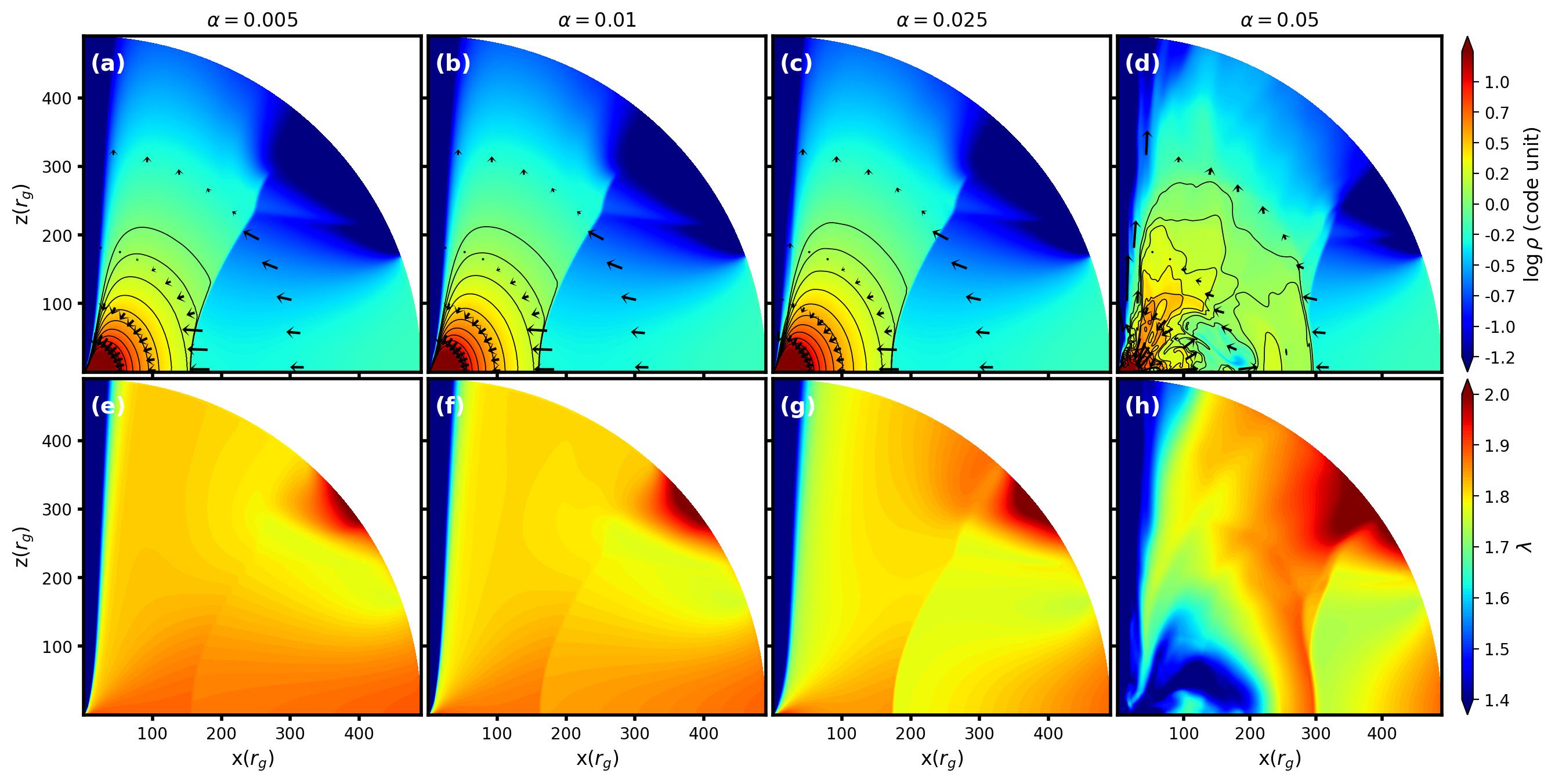}
        \caption{Contours of log$\rho$  overlaid with velocity vectors and $\lambda$, for the model L3. Snapshots (a/e) $\rightarrow$
        (d/h) are captured at times 100000 with $\alpha$ = 0.005, $\alpha$ = 0.01, $\alpha$ = 0.025, $\alpha$ = 0.05, respectively. Ten density contours in logarithmic scale, ranging from $0$ to $1.0$, are shown with black solid lines. See text for details.}
        \label{fig:5}
        \end{center}
\end{figure*}

Figure \ref{fig:5} shows the snapshots of density (top row, panels a-d) at $t=\tin+100000~\tg$ for the model L3 for $\alpha$ = 0.005, 0.01, 0.025, and 0.05. In the bottom row, we plot the corresponding angular momentum distribution (panels e-h). Separate density contour lines were plotted(black) in panels Figs (\ref{fig:5}a-d), which demarcate the shock front (junction of blue and green regions). The shock front is smooth for the steady-state case (panels a-c), but gets perturbed as the post-shock disk becomes time-dependent (panel d). 
Interestingly, despite having higher angular momentum, there is almost no shock oscillation in model  L3 for $\alpha$ = 0.005, 0.01, 0.025. The shock moves outward and disk reaches a steady state, and an accreting torus-like structure is formed in the post-shock disk as shown in Fig. \ref{fig:5}a-c.
The size of this torus increases with the increase of the viscosity parameter $\alpha$. 
Ten contour lines for values of log($\rho$)$=0.0, 0.1,...,1.0$ are over plotted in panels Fig. \ref{fig:5}a-d
, which demarcates the post-shock region. The outer shock is marked out. The inner part of the post-shock disk is like a torus with increasing density. 
Although there is a jump in $\lambda$ across the shock, the angular momentum distribution has mild gradients (Fig. \ref{fig:5}e-f). The angular momentum distribution for $\alpha=0.025$ has a more
pronounced gradient in the pre-shock disk (Fig. \ref{fig:5}g).
Shock oscillation depends on the balance between the outward pressure gradient and the inward gravitational force. The magnitude of the angular momentum of the flow will modify the inward gravitational pull. The higher the angular momentum, the less is the inward pull due to gravity. Higher viscosity implies a much stronger rate of angular momentum transport within the post-shock disk. It reduces the angular momentum distribution closer to the black hole while piling up a lot of $\lambda$ at the shock front, which forces the shock front to shift to a much larger distance from the BH ($\sim 300\rg$ e.g., Fig. \ref{fig:5}d, h). The temperatures close to the BH are larger than those near the shock front. Therefore, the viscous transport of $\lambda$ near the horizon is higher than that near the shock front. It results in a minor piling up of $\lambda$ on the equatorial plane at around a distance of $\sim 100\rg$ (Fig. \ref{fig:5}h). 
Matter flows more smoothly in the region above the equatorial plane and relatively slowly along the equatorial plane. This creates eddies in the post-shock disk; as a result, the torus-like structure within the post-shock disk gets disturbed.



\begin{figure*}
       \includegraphics[width=\textwidth]{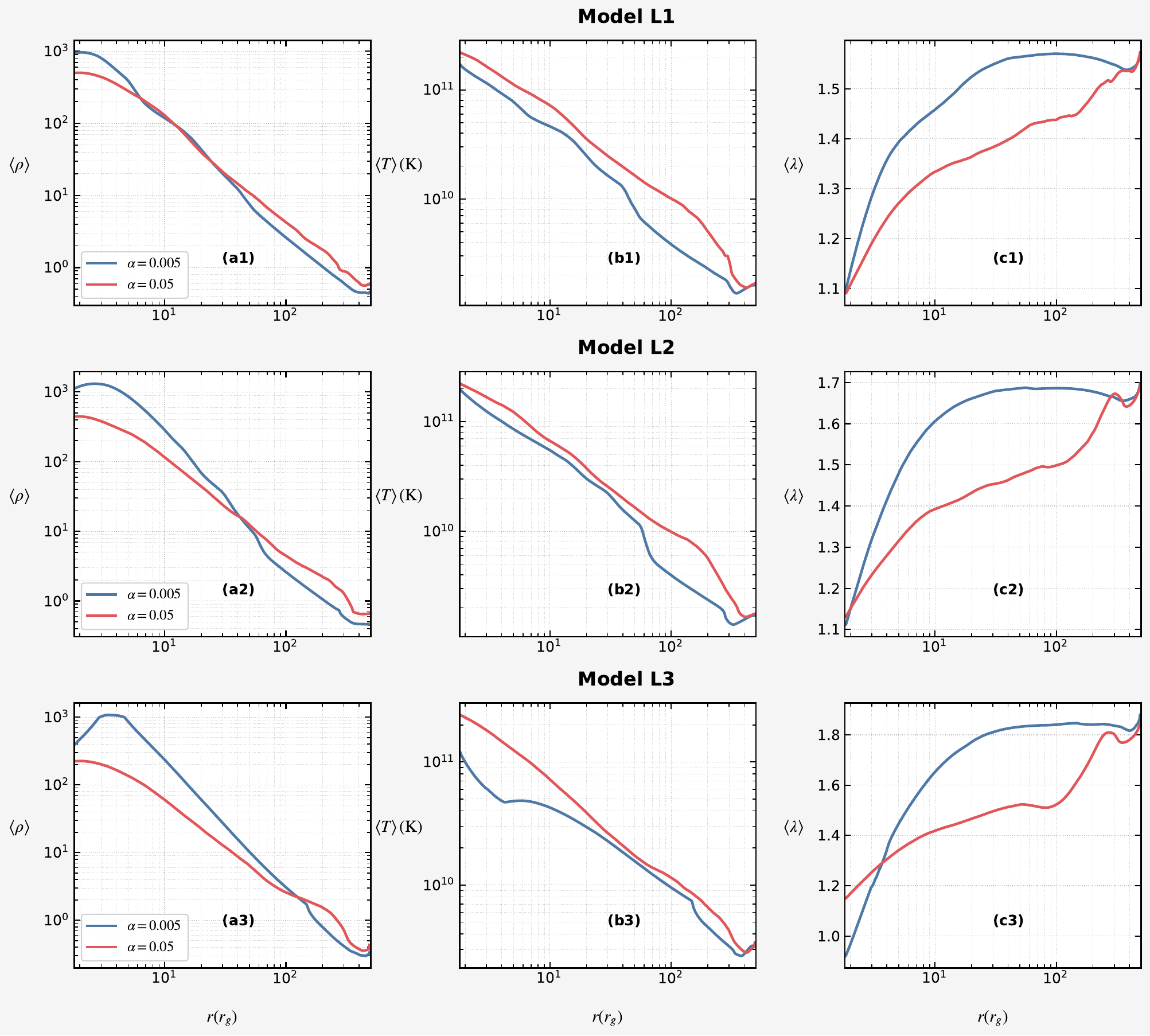}
        \caption{Angle and time-averaged radial distribution of density($<\rho>$), temperature($<T>$) in K, and angular momentum($<\lambda>$). The first, second, and third rows are for models L1, L2, and L3, respectively.}
        \label{fig:6}
\end{figure*}

\begin{figure*}       
\includegraphics[width=\textwidth]{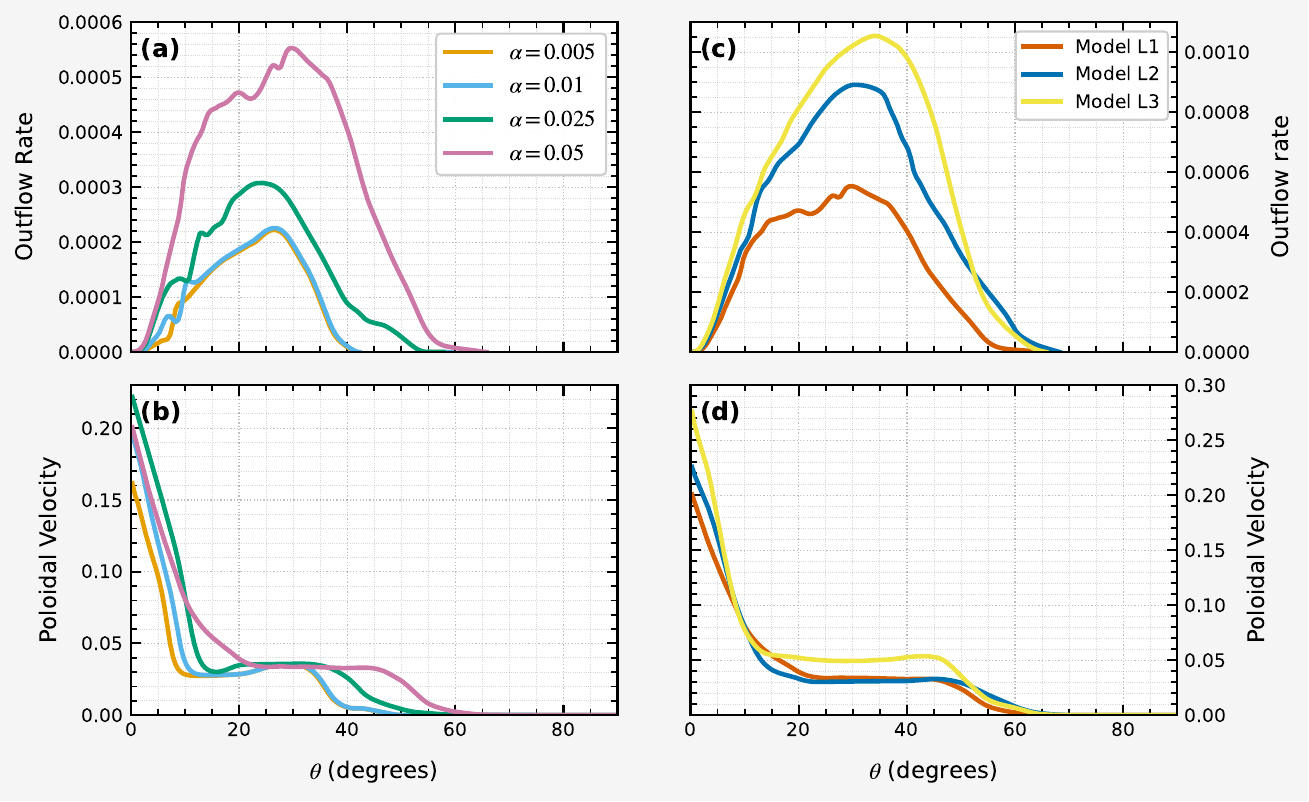}
        \caption{(a) Average outflow mass flux $<\mou >$ vs $\theta$ in units of $\medd$; (b) $<v_p>$ (in unit of $c$) of the outflow measured at $r = 500r_g$ for the model L1, $\alpha$ values as above; (c) $<\mou >$ vs $\theta$, (d) $<v_p>$ vs $\theta$ for $\alpha = 0.05$ of different models.}
        \label{fig:7}
\end{figure*}
\subsection{Angle-averaged accretion profile }\label{subsec:vertical}
In this section, we examine the angle and time-averaged accretion flow variable, e.g., Y, as
\begin{equation} 
\langle Y (r)\rangle  = \frac{1}{N}\sum_n\frac{\sum_j\sin\theta_jY_{nij}\Delta\theta_j}{\sum_j\sin\theta_j \Delta\theta_j }, 
\label{eq:timeav2} 
\end{equation} 
where $Y_{nij} = \rho, T, \lambda$ at time $t=t_n$ and in the cell marked by $r_i$ and $\theta_j$. In this case, the averaging is performed only over cells with $v_r<0$. Over the runtime from \(t_1 = t_{\rm in}\) to \(t_N = t_{\rm in}+100{,}000\,t_g\), we have taken $N=2000$ snapshots, each separated by \(50\,t_g\). Using Eq.\ref{eq:timeav2}, the time \& angle averaged radial profiles of $\langle \rho\rangle$, $<T>$, and $<\lambda>$ of all the three models are presented in Fig. \ref{fig:6}. Each panel compares the average flow variables for two values of the viscosity parameter $\alpha=0.005$ (blue) and $\alpha=0.05$ (red).
In Figs. (\ref{fig:6}a1, a2, a3), at $r\rightarrow$large, $<\rho>$ is higher
for higher values of $\alpha$ (red), but is less at distances closer to the horizon, compared to those due to lower $\alpha$ (blue). While Figs. (\ref{fig:6}b1, b2, b3) show, $<T>$ is higher for flows with higher $\alpha$ at all values of $r$. Flows with less viscosity tend to have a flatter $<\lambda>$ (Figs. \ref{fig:6}c1, c2, c3), only to decrease sharply
at $r<20\rg$. This trend is expected; the angular momentum transport is significant where viscosity is most active. Viscosity is most active where the bulk velocity is low. So in regions where $<\lambda>$ is flat, the velocities are faster, so $<\rho>$ is less. The flow with a higher $\alpha$ has higher viscous dissipation and therefore is hotter.  

From the morphology of the accretion disks
(Figs. \ref{fig:3}, \ref{fig:4} \& \ref{fig:5}), it is clear that the flow patterns are far more complicated than any simple power-law ADAF-like solutions, with multiple time-dependent shocks and bipolar outflows as a natural consequence of the shocks. From Fig. \ref{fig:6}, we show that even time and $\theta$-averaged accretion profiles of transonic, advective disks do not admit a power-law-like description. 

\begin{figure}
    \includegraphics[width=\columnwidth]{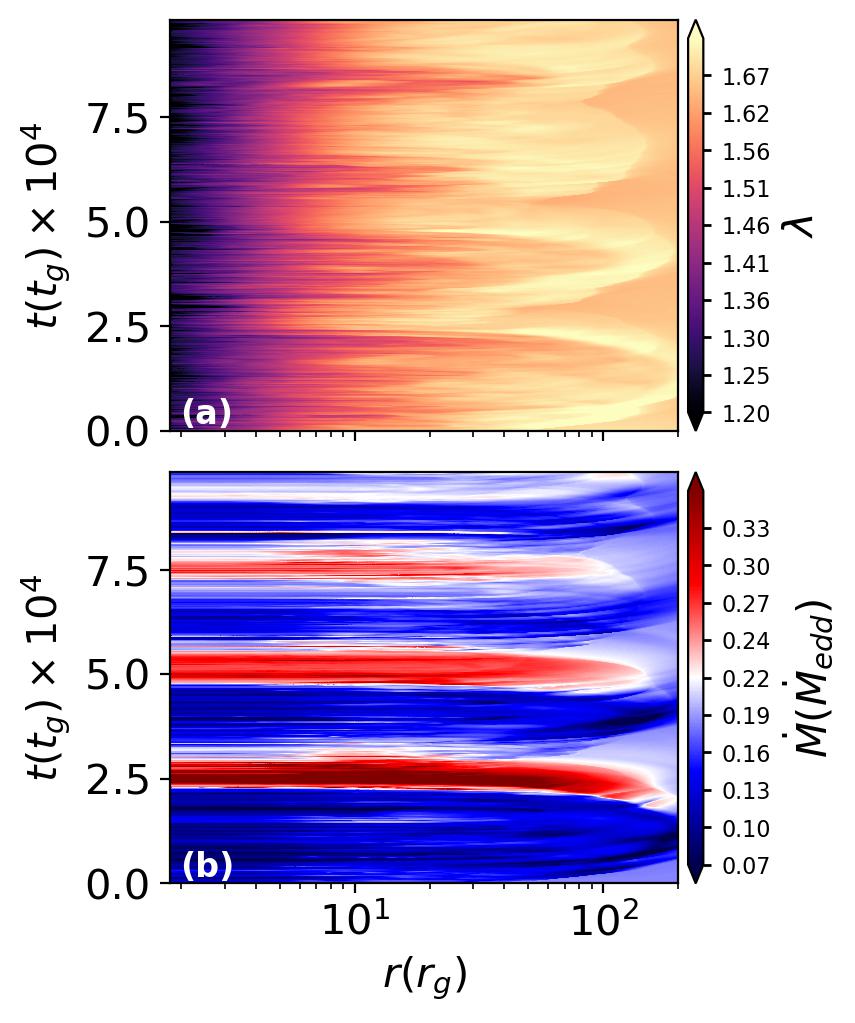}
        \caption{Space-time diagrams showing (a) the angle averaged angular momentum ($<\lambda>$) and (b) the accretion rate ($\dot{M}$ in the units of $\medd$) for viscosity $\alpha$ = 0.025 for the model L2.}
        \label{fig:8}
\end{figure}

\subsection{Average outflow profiles}
From the velocity vectors in Figs. (\ref{fig:1}, \ref{fig:3}, \ref{fig:4}), it is apparent that a fraction of the inflowing matter is deflected as outflows about the axis of symmetry. So we investigate the outflow behavior by studying the average outflow rate and outflow speed at the outer boundary of the computational domain.
We define a time average quantity of a variable $X_{nij}$, 
as
\begin{equation}
\langle X \rangle =\Sigma_{n=1}^N\frac{X_{nij}}{N}
%
\label{eq:timeav}
\end{equation}
{As before, $n,~i,~j$ represents time, radial and $\theta$ indices, respectively. 
In Eq.\ref{eq:timeav}, if we fix the $r$ coordinate, then  
$\langle X \rangle$ become a function of $\theta$.}
We choose $r_i=500\rg$, so $\langle X \rangle$ becomes a function of $\theta$ only at the outer boundary.
In Fig. \ref{fig:7}a, the time-averaged outflow mass flux as a function of $\theta$ is plotted. Here $X_{nij}=2\pi \rho v_r r_i^2 \sin(\theta_j)\, \Delta\theta_j$. 
In Fig. \ref{fig:7}b, the time average of the outflow poloidal velocity is plotted with $\theta$, where
$X_{nij}=v_p= \sqrt{v^2_r+v^2_\theta}$. Both panels (a) \& (b) are for model L1, but for different values of $\alpha$.
The total opening angle of the
outflow increases with the viscosity parameter and is widest for $\alpha=0.05$. Close to the axis of symmetry, where the $\langle \mou\rangle$ is very low, $\langle v_p\rangle \sim 0.2c$ for accretion flows with $\alpha> 0.01$. 
It may be noted that there is a faster, lighter outflow $\langle v_p \rangle>0.15c$ around the axis of symmetry ($<10^o$) and a slower ($v_p \sim 0.025c$), heavier outflow in the cone of $10^o<\theta<45^o$.
Generalizing across all models L1, L2, and L3, we plot the average outflow rate $\langle \mou\rangle$ as a function of $\theta$ in Fig.\ref{fig:7}c for a fixed viscosity of $\alpha = 0.05$, with model L1 shown in red, L2 in blue, and L3 in yellow. The corresponding average poloidal velocity $\langle v_p \rangle$ for each model is presented in Fig. \ref{fig:7}d using the same color scheme.
The mass outflow rate for L3 is almost double that of L1. And close to the axis of symmetry, i.e., $\theta \sim 0$, $\langle v_p\rangle \sim 0.27c$. Even the slow part of the outflow is much faster than that for L1. The opening angle of the outflow increases to $60^o$. It must be noted that $\langle \mou \rangle$ shown in Figs. \ref{fig:7}a \& c flows out from the top half of the accretion disk, a similar amount of mass flows out for $\theta \sim 2\pi/3 \rightarrow \pi$.

 \begin{figure*}
	\begin{center}       \includegraphics[width=\textwidth]{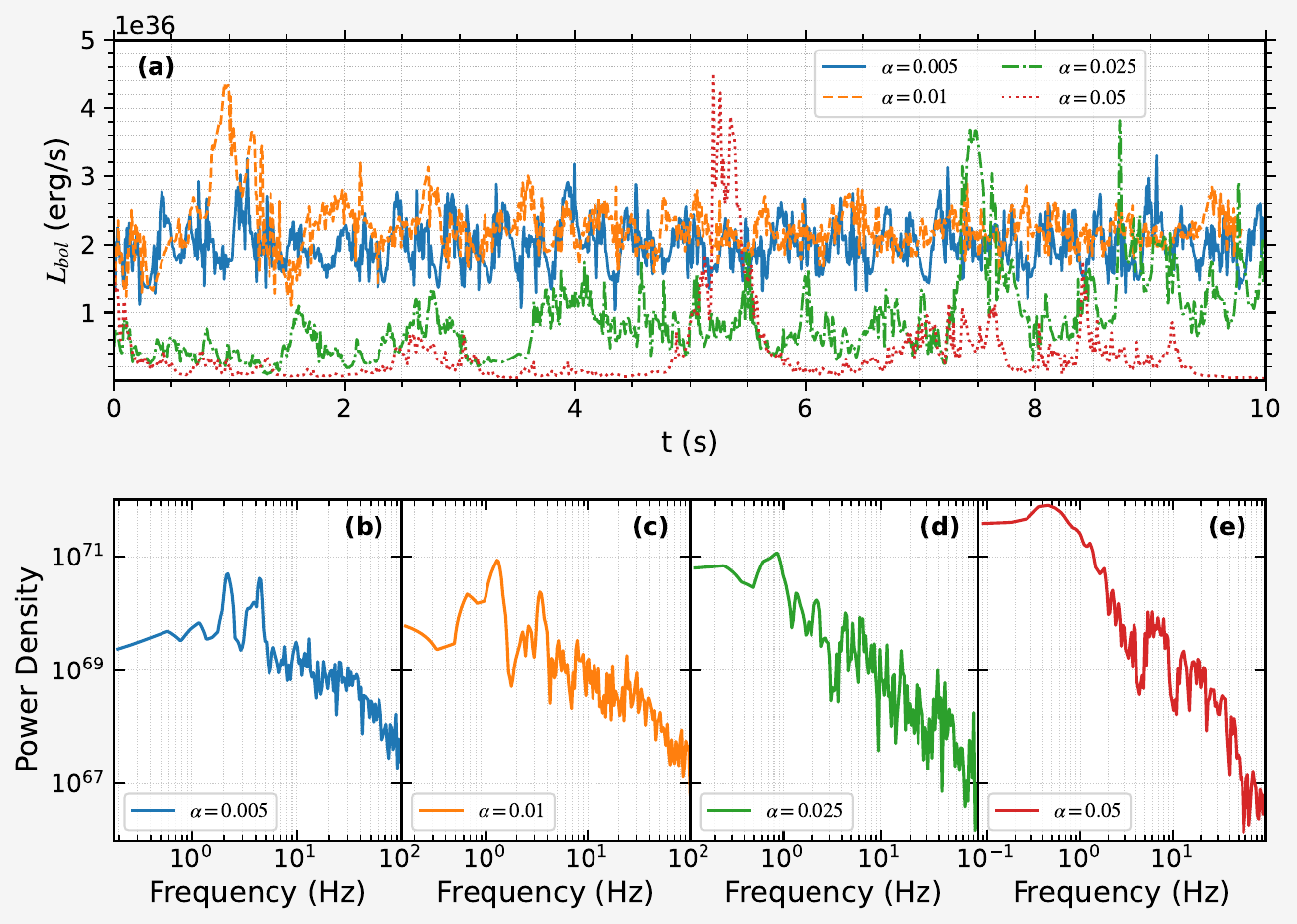}
        \caption{Variation of bolometric luminosity (shown in the upper panel)  in the unit of erg/s and corresponding power density spectrum (shown in the lower panel) with the different $\alpha$ case for the model L2. We consider a black hole mass of $\mbh=10\msol$.}
        \label{fig:9}
        \end{center}
\end{figure*}

\subsection{Time-series Analysis}\label{sec:time_series}
Fig. (\ref{fig:8}a) shows the time evolution of the $\theta$-averaged specific angular momentum $<\lambda>$, and Fig. \ref{fig:8}b shows the mass accretion rate $\dot{M}$ for model L2. 
The viscosity parameter $\alpha=0.025$.
Angular momentum exhibits oscillatory behavior near the shock region with strong amplitude variations as indicated by the darker bands propagating outward. 
This figure shows the persistent long-term inflow, differing from torus-like simulations. Although a constant mass is injected at the outer boundary ($\mdoto=0.3\medd$), the mass accretion rate fluctuates due to shock oscillations, leading to phenomena like quasi-periodic oscillations (QPOs). 
Unlike \cite{2020ApJ...904...21P,2023A&A...678A.141O} who showed QPOs in $\mdoti$ by perturbing shocks by injecting perturbations introduced at the outer boundary, the fluctuations in $\mdoti$ are driven by shock oscillations due to viscosity.

In Fig. (\ref{fig:9}a), time variation of bolometric luminosity ($\lbol$) for the model L2 is plotted for different values of $\alpha$ and by assuming $\mbh=10\msol$. The corresponding power density spectrum is shown in the lower panels of Fig. (\ref{fig:9}b-e). 
Interestingly, $\lbol$ for $\alpha$ = 0.025 and 0.05 has lower values than that with $\alpha$ = 0.005 and 0.01. Although the accretion flow with higher $\alpha$ has higher $<T>$ as shown in Figs. \ref{fig:6}b1, b2, b3, it has a lower $\langle \rho \rangle$ in the inner part of the disk (post-shock disk) as shown in Fig. \ref{fig:6}a1, a2, a3. Cooling is a function of both temperature and density. 
Luminosity additionally depends on the size of the emitter. It has been shown earlier in advective disks that the post-shock disk is the main emitter. In the models discussed in this paper, the shock location increases with $\alpha$. Viscous dissipation also heats the flow, so accretion flows with higher $\alpha$, are hotter (See, Figs. \ref{fig:6}b1, b2, b3; Figs. \ref{fig:3} middle row). So it is expected that the luminosity should increase with the increase in viscosity. However, the luminosity may increase or decrease with the increase of $\alpha$ (see Table \ref{tab:model_properties}). Higher viscosity also means lower angular momentum in the inner disk (see Fig. \ref{fig:6}c2). This makes matter fall into the black hole faster or with higher values of $v_r$. Higher $v_r$ implies lower values of density (Fig. \ref{fig:6}a1, a2, a3) in the inner part of the flow (post-shock region), and emissivity is crucially dependent on $\rho^2$. It also depends on temperature, but the temperature ranges considered here, $T\sim 10^{9-11}$K, are not large enough to dominate emissivity. At high $\alpha$, there is a significant reduction of $\lambda$ in the inner regions (see Figs. \ref{fig:6}c1, c2, c3; \ref{fig:4} middle row), which increases inward $v_r$. Therefore, the density will be significantly lower (Figs. \ref{fig:6}a1, a2, a3) as $\alpha$ increases. The density decreases to the extent that it compensates for the increase in shock location and lowers the luminosity. Therefore, we see a decrease in luminosity in L2 for $\alpha>0.025$. Moreover, there are some spikes in the luminosity curves, e.g., $4.5 < t < 6$s in the red-dot curve. These spikes correspond to the minima of the shock location $\rsh$, at time snap $t\sim 4.4\times10^3$---$6\times10^3~\tg$. As the $\rsh \sim 350 \rightarrow 50 ~\rg$, both the temperature and the density of the post-shock disk increase sharply, giving rise to a spike in luminosity by a factor of 5 (Fig. \ref{fig:9}a). The height of the spikes depended on how much the shock location shifted toward the BH. It may be noted that the $2$s spike inthe red-dot curve is obtained assuming a $10\msol$ BH, but it would correspond to a $1.73$ days spike for a $10^6\msol$ BH, and a $173$ day spike for a $10^8 \msol$ BH. 

The power density spectrum (PDS) for $\mbh=10 \msol$, shows a frequency of 2.2 Hz, and 4.5 for $\alpha$ = 0.005 (Fig. \ref{fig:9}b), 1.3 Hz, and 3.45 for $\alpha$ = 0.01 (Fig. \ref{fig:9}c), 0.85 Hz for $\alpha$ = 0.025 (Fig. \ref{fig:9}d), and 0.44 Hz for $\alpha$ = 0.05 (Fig. \ref{fig:9}e).

\begin{deluxetable*}{lcccccccccBcccc}
\tablewidth{0pt}
\tablecaption{Details of the shock oscillation properties and the average quantities for all three models, the assumed mass of the BH is $10\msol$.} 
\label{tab:model_properties}
\tablehead{
\colhead{Model} & \colhead{$\alpha$} & \colhead{Frequency} & \colhead{Amplitude} & \colhead{Mean $\rsh$}& \colhead{$<\mdoti>$}& \colhead{$<\mou>$} & \colhead{ $\frac{\langle \lbol \rangle}{2}$} \\
 \colhead{}& \colhead{} & \colhead{(Hz)} &\colhead{($r_g$)} & \colhead{($r_g$)} & \colhead{($\medd$)} & \colhead{($\medd$)} &  \colhead{(erg/s)$\times 10^{35}$}}

\startdata
L1 & 0.005 & 2.21 & 6.76 & 47.22 & 0.179 & 0.018 & 5.87\\
 & 0.01  & 1.36 & 9.60 & 55.69 & 0.175 & 0.019 & 8.36 \\
   & 0.025 & 0.50 & 76.06 & 124.41 & 0.168 & 0.029 & 8.73 \\
   & 0.05  & 0.24 & 178.55 & 282.53 & 0.116 & 0.071 & 5.49 \\
\hline
L2 & 0.005 & 2.04 & 5.67 & 60.51 & 0.181 & 0.018 & 19.88 \\
   & 0.01  & 0.72 & 6.84 & 68.44 & 0.180 & 0.018 & 22.12 \\
   & 0.025 & 0.31 & 71.52 & 139.66 & 0.156 & 0.036 & 24.96 \\
   & 0.05  & 0.16 & 113.47 & 317.71 & 0.088 & 0.110 & 3.94 \\
\hline
 L3 & 0.005 & - & - &156.62 & 0.050 & 0.040 & 15.42 \\
   & 0.01  & - & - & 159.22 & 0.053 & 0.040 & 14.91 \\
   & 0.025 & - & - & 190.76 & 0.060 & 0.051 & 10.62 \\
   & 0.05  & 0.33 & 71.90 & 283.83 & 0.042 & 0.137 & 1.85 \\   
\enddata
\end{deluxetable*}

Table \ref{tab:model_properties} presents the 
$\nu_{\rm QPO}$, amplitude of the shock oscillation, mean shock location, $\langle \mdoti \rangle$, $\langle \mou \rangle$ and $\langle \lbol/2 \rangle$ for different values of $\alpha$ for all the three models. For models L1 and L2, shock oscillates for all four $\alpha$, L3 has a threshold $\alpha$ ($>0.025$) to oscillate the disk. The mean position and the amplitude of the oscillating shock increase with the increase in $\alpha$, while the oscillation frequency decreases. This behavior is expected; increasing the mean shock position with $\alpha$ will lead to a longer oscillation period.  
For the same $\alpha$, but with different models, an increase in angular momentum results in a larger mean position and amplitude of oscillation, accompanied by a decrease in $\nu_{\rm QPO}$. \cite{2024MNRAS.528.3964D} had similar conclusions, although the radiative loss was computed a posteriori.  
Additionally, $<\mdoti>$ decreases but $<\mou>$ increases  with the increase in $\alpha$. 
It may be noted that eddies triggered by different rates of $\lambda$ transport along different $\theta$, inhibit mass supply to the BH and enhance the fraction of mass leaving the domain. In Table \ref{tab:model_properties}, in models L1 \& L2, $\lbol$ do not monotonically decrease with increase in $\alpha$. For lower values of $\alpha$, the angular momentum transport in the post-shock disk is insignificant. Therefore, the increase in $v_r$ or decreases in $\rho$ is not significant enough to compensate for the increase in the size of the main emitter, or the post-shock disk. So, for smaller values of $\alpha$, $\lbol$ increases with $\alpha$, but the trend reverses for higher values.

\section{Summary And discussion}\label{sec:summary}
In this work, we perform hydrodynamical simulations with a second-order accurate code in both space and time to investigate the effects of viscosity on accretion disks in the presence of cooling. 
We have chosen three models: L1, L2, and L3, with different angular momentum at the outer boundary. 
We consider a constant mass supply of 0.3$\medd$ at the outer boundary, which allows us to study the long-term evolution of the accretion disk. Ours is a 2D simulation, while the real world is three-dimensional (3D). \citet{2023MNRAS.519.4550G} reported that the dynamical vortices observed in 2D simulations retain their axisymmetry when extended to 3D. To deviate from 2D, non-axisymmetric processes need to be considered. This paper stresses how viscosity in the presence of cooling might affect outflow formation and QPO generation and their interconnection. If we retain similar boundary conditions, a 3D simulation would also produce similar accretion disk properties.
The key highlights of our paper are:
\begin{enumerate}
    \item Semi-analytical solutions of models L1 and L3, do not admit accretion shocks. However, numerical simulations admit shocks. Therefore, the shock parameter space broadens in two dimensions. In multidimensional flows, accretion occurs from off-equatorial planes, as well as along the equator. Even if the shock conditions are not satisfied on the equatorial plane (as in analytical solutions), they may still be met due to the intervention from the off-equatorial matter flow (see Figs. \ref{fig:2}, \ref{fig:3}, \ref{fig:4}) 
    This is also the fundamental difference between 1D and 2D simulations.
    \item Viscosity redistributes angular momentum and heats the flow, which may destabilize shocks. Incorporation of cooling stabilizes the shock.
    \item Viscous angular momentum transport varies with angle $\theta$, which produces local eddies in the post-shock disk, leading to the formation of hot bubbles. 
    The mass accretion rate thwarts the outward motion of bubbles along the equatorial plane, but those along lower $\theta$ leave the domain along the outflow.
    \item Post-shock disk produces fast outflows ($\langle v_{\rm p} \rangle \sim 0.27c$) close to the axis of symmetry, and slower ($\langle v_{\rm p}\rangle > 0.03c$) at wider angles. These are time averages, so at times the outflow speed can be even faster.
    \item The profiles of transonic accretion are not self-similar. Angle and time averaging smoothens the accretion profiles and reduces the sharp jumps due to shocks, yet transonic accretion solutions cannot be matched with simple power-law profiles.  
    \item Time variation of bolometric luminosity reveals quasi-periodic oscillations which closely match the shock oscillation. The QPO frequencies for $10\msol$ BH ranged from a few Hz to sub-Hz. The luminosity is low because not all possible cooling mechanisms are included \citep{2020A&A...642A.209S,2023MNRAS.522.3735S}. 
\end{enumerate}

\begin{acknowledgments}
The authors acknowledge the Surya computer cluster facility hosted by ARIES, Nainital, India, and the DANTE platform, APC, Paris, France. The authors also acknowledge the anonymous referee for fruitful suggestions.
\end{acknowledgments}



\bibliography{biblio}{}
\bibliographystyle{aasjournal}



\end{document}